\documentclass[nonacm,screen]{acmart}
\thanks{CSCW 2021 Workshop -- Investigating and Mitigating Biases in Crowdsourced Data, October 23, 2021, Virtual.\\ \copyright 2021 Copyright held by the author(s).}
\authorsaddresses{}
\AtBeginDocument{%
  \providecommand\BibTeX{{%
    \normalfont B\kern-0.5em{\scshape i\kern-0.25em b}\kern-0.8em\TeX}}}

\setcopyright{acmcopyright}
\copyrightyear{2021}
\acmYear{2021}
\acmDOI{10.1145/1122445.1122456}

\acmConference[BCD21]{Investigating and Mitigating Biases in Crowdsourced Data Workshop at ACM CSCW 2021}{October 24, 2021}{Virtual}

\usepackage{xcolor} 
\usepackage[nomessages]{fp} 
\usepackage{booktabs} 
\usepackage{multirow} 
\usepackage{url}
\usepackage{framed}
\usepackage{enumitem}
\usepackage{array}
\usepackage{tabularx}
\usepackage{balance}
\usepackage{amsmath} 
\usepackage{siunitx} 
\usepackage{subcaption}
\usepackage{adjustbox}
\usepackage[flushleft]{threeparttable}
\usepackage{setspace} 
\usepackage{caption} 

\definecolor{ForestGreen}{rgb}{0.13, 0.55, 0.13}
\definecolor{tomBlue}{HTML}{0196CE}

\usepackage{soul,color}
\soulregister\cite7
\soulregister\ref7
\soulregister\pageref7
\setstcolor{red}
\setul{}{.2ex}

\newcommand{\sm}[1]{\textcolor{red}{[{\bf SM: #1}]}}

\newcommand{\kr}[1]{\textcolor{brown}{[{\bf KR: #1}]}}

\usepackage{xspace}

\begin{document}

\title{Bias Management in Crowdsourced Data: Moving Beyond Bias Mitigation and\sm{non parliamo di "Mitigation", lo toglierei.}  Removal \sm{a me piacerebbe di piu' "From Bias Removal to Bias Management", o "Bias Management: Beyond Bias Removal [in Crowdsourced Data]"}
\kr{Bias Management in Crowdsourced Data: Moving Beyond Bias Removal.}
}

\title{Managing Bias in Human-Annotated Data: Moving Beyond Bias Removal}


\author{Gianluca Demartini}
  \affiliation{
  \institution{The University of Queensland, Brisbane, Australia}
 }
\author{Kevin Roitero}
  \affiliation{
  \institution{University of Udine, Udine, Italy}
 }
 \author{Stefano Mizzaro}
  \affiliation{
  \institution{University of Udine, Udine, Italy}
 }


\begin{abstract}
 Due to the widespread use of data-powered systems in our everyday lives, the notions of bias and fairness gained significant attention among researchers and practitioners, in both  industry and academia. Such issues 
 typically emerge
 from   the data, which comes with varying levels of quality, used to train systems.
 With the commercialization and employment of such systems that are sometimes delegated to make life-changing decisions, a significant effort is being made towards the identification and removal of possible  sources of bias that may surface to the final end-user.
 In this position paper, we instead argue that bias is not something that should necessarily be removed in all cases, and the attention and effort should shift from bias removal to the identification, measurement, indexing, surfacing, and adjustment  of bias, which we name \emph{bias management}.
 We argue that if correctly managed, bias can be a  resource that can be made transparent to the the users and empower them to make informed choices about their experience with the system.
\end{abstract}

\maketitle

\section{To Remove Bias or Not To Remove Bias: That is The Question}

Different humans have diverse experiences and backgrounds which lead
to them having different points of views. Each having a subjective view of the world, behavioral sciences agreed 
that humans are subject to systematic biases and errors \cite{kruglanski1983bias}.
Our society has been delegating more and more tasks and decisions to data-driven algorithms and to automatic or semi-automatic computational systems.
Despite efforts to keep the algorithms behind data-driven computational systems neutral and unbiased, joint with the fact that such systems are often trained on human-annotated data, popular incidents made people realize that algorithms and datasets are not free from biases \cite{danks2017algorithmic,tommasi2017deeper}. 
Famous examples include the case where an algorithm designed to predict the likelihood of a criminal offending was found to be racially biased, and, according to the system, black people where predicted to have a higher risk of recidivism then their true one, and the reverse for white people \cite{compas-recidivism}.
Another example is the study which found that facial recognition technology software used for law enforcement was correct 99\% of the times for white men, while for dark-skinned women the accuracy was  less than 35\% \cite{face-bias}. Another study showed that the search results of a search engine for the keyword ``CEO'' and for highly paid jobs were gender biased \cite{datta2015automated,otterbacher2018investigating}. 
As we can also see with the recent case of Facebook needing to apologize as black men were labeled `primates'\footnote{``Facebook apology as AI labels black men `primates' ''. 7 Sep 2021. \url{https://www.bbc.com/news/technology-58462511}}, even large internet platforms are still prone to such bias-driven mistakes.

To address these issues, the study of bias in data and  algorithms gained popularity \cite{kirkpatrick2016battling,tommasi2017deeper,baeza2018bias,obermeyer2019dissecting,xu2020investigating,saleiro2020dealing}, both in disciplines that study fully-automatic systems such as machine and deep learning \cite{noseworthy2020assessing,mehrabi2021survey} and in those that study and develop 
crowdsourcing-based or hybrid human-in-the-loop systems \cite{faltings2014incentives,eickhoff2018cognitive,demartini2020human,demartini2017introduction}.
Analyzing the recent literature, a clear trend that emerges is that algorithmic and human biases are depicted as negative and undesirable, and that bias should be removed from data and systems thus enforcing a, perhaps utopian, completely un-biased  system and output  \cite{kearns2019empirical,jung2019eliciting,oneto2020fairness}.

In this position paper, we argue that
(i) bias removal could be harmful, as it can lead to users being presented with a reality which is different from that in the off-line world, and 
(ii) if properly managed, bias can be not only harmless but even useful, as it could be a valuable source of information for the end user.
Therefore, we claim that the aim should not be to remove bias a-priori; instead, bias should be identified, measured, indexed, surfaced to users, and treated as a feature of the system, delegating to the users the choice of whether and how to adjust for it.
Note that item (i) is consistent with Ullman's view that ``we should not blame data if it reflects the world as it is, rather than as we would like it to be'' \cite{ullman2020battle}, but item (ii) goes beyond that.

\section{Bias Management vs. Bias Removal}

We present two use cases for which we envision a scenario where bias removal may be undesirable and potentially harmful and where we think that, instead, bias management would be a more effective approach.

\subsection{Example 1: Search Engine}

Consider the case where a user needs to manually annotate or label a set of images to create a dataset; it is reasonable to assume that such a dataset may then be used to train an automatic system to independently perform a specific task.
Suppose that the user issues the gender-neutral query ``nurse'' on a search engine and searches for images. The user will see on the page of results the vast majority of images of female nurses. 
While this might appear as an indication that the ranking algorithm of the search engine has a gender bias, this might also reflect the real gender distribution in this profession, that is, for example, female nurses are statistically more frequent than male nurses. 
While a traditional approach might propose to resolve this bias by forcing the algorithm to show male and female nurses in the same percentage, we argue that an alternative, less invasive approach might be more useful to the user. We make the following proposal to address the issue in this use case.
The search engine might display on the result page a set of additional metadata which may be useful to the user to have a complete understanding of the magnitude of bias in the result set; for example, the search engine might show a label indicating that ``the search results appear to be highly imbalanced in terms of gender: in the top 1000 results, 870 of them are of female nurses and 130 of them are of male nurses''. This information makes the user informed and aware of the statistical distribution of the search results with respect to a specific group (in this case, gender). Then, ideally, the user should be asked if they would like to maintain the current result ranking or whether they would prefer to inspect the results after a fairness policy is applied to the data (in this case, for example, forcing the number of male and female nurses to be roughly the same in the search result list). 

We argue that not employing an explicit and transparent bias removal intervention might be potentially harmful to the user. In fact, if the task of the user was to investigate something related to or influenced by the percentage of male and female nurses, the implicit application of the fairness policy might leave the user with an inaccurate perception of the real gender distribution in the nursing profession. Taking this concept to the extreme, the user might even erroneously think, somehow paradoxically, that gender bias is not present in the nursing profession, and that male and female nurses are equally present in the job market.

\subsection{Example 2: Recommender System}
Consider a recommender system with a highly unbalanced set of users; for the sake of simplicity, let us suppose that there are just two groups of users: A, which constitutes 90\% of the system userbase, and B, which constitutes the remaining 10\% of the userbase (e.g., A could correspond to male  and B to female users). Let us also suppose that A and B have very different tastes, and that a product which is good for users belonging to group A is generally not appreciated by users belonging to group B (e.g., the product could be a movie).
Let us suppose that the recommender  system in production is trained to maximize for user engagement. Given the unbalanced userbase composition, the recommender system will probably serve most of the times items that are likely to be enjoyed by group A users and disliked by users from group B (in an ideal case, the system should learn to recommend different items to users in A and B, but let us also assume that this cannot be done, for example for privacy/anonymity reasons or to remove group belonging information from the model for fairness purposes).
In this case, enforcing a fairness policy would mean to serve 50\% A-liked items and 50\% B-liked items. However, this might be extremely harmful: the risk is that many users will be not satisfied with the product being served. 
Notice that if the majority of the users (in this case, group A) is served with an item that they do not like, this will result in a loss of effectiveness of the whole recommender system. 
On the contrary, we argue that in this case either the users should have a clear overview of the rationale behind being recommended with a particular item (i.e., an explainable system), or the designers of the recommender system should identify a fairness-effectiveness trade-off.
Note that imbalanced labeled datasets like the one presented in this use case commonly lead to \textit{unknown unknown} errors, that is, to trained models that result in high-confidence errors. These errors are very difficult to identify as the model is highly confident of having made an accurate classification decision \cite{dong2020region}.

\section{The Bias Management Workflow}

The two examples described above serve to support our position that the answer to the question ``should bias always be removed and fairness always enforced'' is not as straightforward as it might seem at a first glance. 
Our proposal alternative to removal consists of different steps, which are detailed in the following. 

\begin{enumerate}
    \item \emph{Identification}: identify if the data or system being used is subject to bias or fairness issues.
    \item \emph{Measurement}: quantify with an appropriate metric the magnitude of different types of bias present in the data or system which is under consideration.
    \item \emph{Indexing}: collect, parse, structure, and store bias metadata and fairness policies aimed at facilitating a subsequent fast and effective retrieval and system adjustments. 
    \item \emph{Surfacing}: present in an appropriate way to the user the bias present in the underlying data and/or any fairness policy that have been applied to the  data or system  under consideration.
    \item \emph{Adjustment}: provide the user with a set of tools which allows them to interact with existing bias and to adjust for it in their preferred ways. This enables them to make  informed decisions. Giving control to the user is essential since for some tasks they may benefit from fairness (e.g., a job application scenario) while for some others they may not (e.g., understanding the gender distribution in a specific profession). 
\end{enumerate}

We argue that bias is part of human nature, and that it should be managed rather than removed. 
%
We believe that the ideas detailed in this position paper can lead to a more sound, informed, and transparent decision making process which will impact algorithm and system design.
%

In the future we plan to categorize use cases across different domains,  to design a \emph{bias management system} implementing (maybe with a human-in-the-loop approach) the above five-step pipeline, and to evaluate the impact of such an approach in terms of effectiveness, user satisfaction, and user engagement.

\newpage
\bibliographystyle{ACM-Reference-Format}
\bibliography{paper}

\end{document}